\documentstyle[12pt]{article}
\def\a{\alpha}
\def\b{\beta}

\def\l{\lambda}
\def\m{\mu}
\def\n{\nu}

\def\d{\delta}
\def\e{\epsilon}
\def\t{\theta}
\def\be{\begin{equation}}
\def\ee{\end{equation}}
\def\p{\partial}
\def\ber{\begin{eqnarray}}
\def\eer{\end{eqnarray}}
\begin{document}

\begin{center}
{\Large\bf Noncommuting Electric Fields and
Algebraic Consistency in Noncommutative Gauge Theories}
\vskip 1 true cm
{\bf Rabin Banerjee}\footnote{On leave from S.N.Bose Natl. Ctr. for Basic Sciences,
Calcutta, India; e-mail:rabin@post.kek.jp;  rabin@bose.res.in}
\vskip .8 true cm
Institute of particle and nuclear studies,

High Energy Accelerator Research Organisation (KEK),

  Tsukuba 305-0801, Japan
\end{center}

\bigskip

\centerline{\large\bf Abstract}
\medskip

We show that noncommuting electric fields occur naturally in $\t$-expanded
noncommutative gauge theories. Using this noncommutativity, which is field
dependent, and a  hamiltonian generalisation of the Seiberg-Witten map,
the algebraic consistency in the lagrangian and
hamiltonain formulations of these theories, is established. A comparison of results in different descriptions shows that this generalised map acts as a canonical transformation in the physical subspace only. Finally, we apply the hamiltonian formulation to derive the gauge symmetries of the action.

\newpage

\section{Introduction}

Snyder's \cite{s} old idea that spatial coordinates
do not commute has undergone a recent revival due to its appearance in
string theory \cite{sw}. Inspired by this fact, several papers have
appeared discussing different aspects of quantum mechanics and field
theory on noncommutative space \cite{r}. 

There are, however, some issues which have received less attention than
others. For instance, only a few papers \cite{d, g, af, ab}
discuss  the hamiltonian
treatment of noncommutative gauge theories. Indeed   one possible approach
to study these theories is to exploit the Seiberg-Witten map
\cite{sw} that yields a commutative equivalent to the original
noncommutative theory. However this transition is done at the level of
the lagrangian. Also, nonabelian theories \cite{af, js} have not been that
widely studied as their abelian counterparts. The issue of algebraic
consistency among the various approaches is left open. Likewise, the implications of the Seiberg-Witten map in constructing effective theories from the noncommutative theories is not completely clear. Indeed it is known that the energy -momentum tensor obtained directly from the noncommutative theory and then using the map is different from the expression obtained by considering the action of the effective theory \cite{gkp}.  

In this paper we show that noncommuting electric fields occur in
noncommutative gauge theories, mapped to their commutative equivalents by
the Seiberg-Witten transformation. This algebra is used to obtain the
equations of motion from the hamiltonian.Its equivalence with the
lagrangian equations of motion is established. By a suitable
redefinition, we show that both the lagrangian and hamiltonian can be put
in the usual form as the difference or the sum of the squares of the
electric and magnetic fields. The entire effect of noncommutativity is
shifted to the nontrivial algebra among these fields.
A hamiltonian generalisation of
the Seiberg-Witten map is obtained. This is used to show that the
commutative equivalents obtained either from the noncommutative
lagrangian or hamiltonian are compatible. In conformity with \cite{gkp} we find that the computation of the hamiltonian density does not commute with the Seiberg-Witten expansion and the star products. The result is different depending upon whether it is obtained directly from the effective theory or whether it is obtained in the noncommutative version, after which the map is exploited. However we find that, after implementing the Gauss constraint, this difference is a total boundary, so that the expressions for the hamiltonian agree. The stability of the
Poisson algebra among the electric and magnetic fields, under the generalised Seiberg-Witten 
map, is examined.  This is used to clarify certain issues regarding the possible interpretation of  this map as a canonical transformation. Our analysis is for Yang-Mills theory with
$U(N)$ gauge group, including, in particular $N=1$ (i.e. Maxwell's
theory).

An application of the hamiltonian analysis has been discussed in details where the gauge symmetries of the action are systematically derived, following the Dirac algorithm. For the $U(1)$ case it implies that the equations of motion can always be put in a `Maxwell'-like form.

After setting up the notations, we carry out the hamiltonian analysis, both in the noncommutative and effective theories, in section 2.  Section 3 contains our analysis of the Seiberg-Witten map as a canonical transformation. The hamiltonian formulation is used to derive, in section 4,  the gauge symmetries of the action in the noncommutative variables, while the concluding remarks are given in section 5.

The ordinary Yang-Mills action is given by,
\begin{equation}
S=-\frac{1}{4}\int d^4x \,\,Tr(F_{\m\n}F^{\m\n})
\label{1}
\end{equation}
where the nonabelian field strength is defined as usual,
\be
F_{\m\n}=\p_\m A_\n -\p_\n A_\m -i[A_\m, A_\n]
\ee
\label{2}
in terms of the gluon field,
\be
A_\m (x)= A_\m^a(x) T^a
\label{3}
\ee
Here $T^a$ are the generators of a $U(N)$ gauge group satisfying,
\be
[T^a, T^b]=if^{abc}T^c, \,\, \{T^a, T^b\}=d^{abc}T^c,\,\, Tr(T^aT^b)
=\d^{ab}
\label{4}
\ee

The noncommutative generalisation of this theory involves the star 
product of the noncommutative field strength $\hat F_{\m\n}$, 
expressed in terms of the field $\hat A_\m$,
\be
\hat F_{\m\n}=\p_\m \hat A_\n -\p_\n \hat A_\m -i(\hat A_\m * \hat A_\n
- \hat A_\n * \hat A_\m)
\label{5}
\ee
so that,
\be
\hat S=-\frac{1}{4}\int d^4x \,\,Tr(\hat F_{\m\n}*\hat F^{\m\n})
= -\frac{1}{4}\int d^4x \,\,Tr(\hat F_{\m\n}\hat F^{\m\n})
\label{6}
\ee
where the second equality follows on using the definition of the star
product,
\be
(A * B)(x) = e^{\frac{i}{2}\t^{\a\b}\p_\a\p'_\b} A(x)B(x')|_{x'=x}
\label{7}
\ee
and dropping boundary terms. Here $\t^{\m\n}$ is a real and antisymmetric
constant matrix.

\section{The hamiltonian analysis}

We now carry out a hamiltonian analysis
of this theory. In order to avoid higher order time  derivatives ,
henceforth $\t^{\a\b}$ will be chosen to have only spatial components so
that $\t^{0\a}=0$ and $\t^{ij}=\e^{ijk}\t^k$. Also, the ensuing
analysis will be confined to the leading order in $\t$ only. First, the effective theory obtained by the Seiberg-Witten map \cite{sw}  is considered.

\subsection{The effective theory}

To first order in $\t$ it is possible to relate the variables in the
noncommutative spacetime with those in the usual one by the map
\cite{sw},
\be
\hat A_\m = A_\m -\frac{1}{4}\t^{\a\b}\{A_\a , \p_\b A_\m +
F_{\b\m}\}+O(\t^2)
\label{8}
\ee
\be
\hat F_{\m\n}= F_{\m\n}+\frac{1}{4}\t^{\a\b}(2\{F_{\m\a}, F_{\n\b}\}
- \{A_\a, D_\b F_{\m\n}+\p_\b F_{\m\n}\})+O(\t^2)
\label{9}
\ee
where the covariant derivative is defined as,
\be
D_\m\l =\p_\m\l + i[\l, A_\m]
\label{10}
\ee

On applying this map, the action (\ref{6}) is written in terms of an
effective theory comprising the usual variables \cite{bgp},
\be
\hat S\rightarrow S_{eff}=-{1\over 4}\int d^4x Tr
\Big( F_{\m\n} F^{\m\n}+{1\over 2}
\t^{\a\b}\Big(2\{F_{\m\a}, 
F_{\n\b}\}-
 \{A_\a,  D_\b F_{\m\n} +\p_\b F_{\m\n}\}\Big)
F^{\m\n}\Big)+O(\t^2)
\label{11}
\ee
The above form is further simplified by dropping a boundary term that
does not affect the equations of motion, to yield the following
lagrangian, expressed solely in terms of the field strength,
\be
{\cal L}_{eff}= -{1\over 4} \,\,Tr
\Big( F_{\m\n} F^{\m\n}+{1\over 2}
\t^{\a\b}\Big(2\{F_{\m\a}, 
F_{\n\b}\}+{1\over 2} \{F_{\b\a}, F_{\m\n}\}\Big)
F^{\m\n}\Big)+O(\t^2)
\label{12}
\ee
In this form the gauge invariance of the theory under the usual gauge
transformations $ (\d A_\m = D_\m \a)$ becomes manifest.

Since the
lagrangian is written in terms of the field strengths, it is possible to
work with the electric and magnetic fields, exactly as happened in the
case of the abelian theory
\cite{gjp},
\be
E_i^a= -F_{0i}^a, \,\,\, B_i^a={1\over 2}\e_{ijk}F_{jk}^a
\label{13}
\ee
 
In terms of these variables, the effective lagrangian (\ref{12}) becomes,
\be
{\cal L}_{eff}= {1\over 2}\,\, Tr\Big({(\bf E}^2-{\bf B}^2)
(1-{\bf \t .
B})  + {\bf \t . E} ({\bf B.E} + {\bf E.B})\Big)
\label{13.1}
\ee

The canonical momenta $\pi_\m^a$, conjugate to $A^{\m a}$, are found to
be,
\be
\pi_0^a = 0
\label{14}
\ee
which is the primary constraint of the theory, while the true momenta
are,
\be
\pi_i^a=E_i^a +{1\over 2}\,\,Tr\Big[\Big(\{B_i, \t^j E_j\} - \{E_i, \t^j
B_j\} + \t^i\{B_j, E_j\}\Big)
T^a\Big]
\label{15}
\ee

Since the canonical variables are $(A_i, \pi_i)$, it is useful for later
convenience to invert the above relation and solve the electric field in
terms of the momenta,
\be
E_i^a=\pi_i^a -{1\over 2}\,\,Tr\Big[\Big(\{B_i, \t^j \pi_j\} - \{\pi_i,
\t^j B_j\} + \t^i\{B_j, \pi_j\}\Big)
T^a\Big]
\label{16}
\ee

The hamiltonian is now obtained in the usual way by a Legendre transform,
\be
H_{eff}= \int d^3x (-\pi_i^a \dot A_i^a - {\cal L}_{eff})
   =\int d^3x (\pi_i^a E_i^a +A_0^a(D_i\pi_i)^a - {\cal L}_{eff})
\label{17}
\ee

Time conservation of the primary constraint yields a secondary
constraint. For that it is essential to express the hamiltonian in terms
of the canonical variables. In this case it is obvious from (\ref{13.1})
and (\ref{16}) that the dependance of the hamiltonian on $A_0$ has been
isolated completely. The secondary constraint is therefore the usual
Gauss constraint,
\be
(D_i\pi_i)^a = 0
\label{18}
\ee

It is also possible to verify that no further constraints are generated
by this iterative prescription. As expected, there are only first class
constraints in the theory, which annihilate the physical states of the
theory. The physical hamiltonian; i.e. the hamiltonian acting on the
physical states, is thus given by dropping the second term of the last
expression in (\ref{17}). Using (\ref{13.1}) and (\ref{15})  in
(\ref{17}), we get,
\be
H_{eff}= {1\over 2}\,\,\int d^3x\,\, Tr\Big({(\bf E}^2+{\bf B}^2)
(1-{\bf \t .
B})  + {\bf \t . E} ({\bf B.E} + {\bf E.B})\Big)
\label{19}
\ee
which has a very similar structure as the effective lagrangian
(\ref{13.1}). Note also that setting the noncommutative parameter to zero
reproduces the standard expression for the Yang-Mills theory. For the
abelian theory, the above form was obtained and discussed in \cite{k}.

The positive definiteness of the above hamiltonian is not manifest. It is
however possible to redefine the electric and magnetic fields,
\be
\bar E_i^a = E_i^a -{1\over 4}\t^j d^{abc}(B_j^b E_i^c -2 E_j^b B_i^c)
\label{19a}
\ee
\be
\bar B_i^a = B_i^a -{1\over 4}\t^j d^{abc} B_j^b B_i^c
\label{19b}
\ee
so that the hamiltonian, in these variables, becomes,
\be
H_{eff}= {1\over 2}\,\,\int d^3x\,\, Tr( {\bf \bar E}^2+{\bf \bar B}^2)
\label{19c}
\ee
It is now structurally identical to the hamiltonian for the ordinary
Yang-Mills theory. The entire effect of noncommutativity will be shifted
to the nontrivial algebra among the electric and magnetic fields, a point
which we shall consider in some details later. Also note that in these
variables, the lagrangian (\ref{13.1}) has the form of the usual
Yang-Mills lagrangian,
\be
{\cal L}_{eff}= {1\over 2}\,\, Tr{(\bf \bar E}^2-{\bf \bar B}^2)
\label{19d}
\ee

\subsection{Analysis in the noncommutative variables}

We shall now consider the lagrangian following from the action (\ref{6})
in the original hat variables and obtain the  hamiltonian. This will be expressed in terms of the usual variables by using the
 map (\ref{9}), so that a comparison can be made with (\ref{19}). We find that although the hamiltonian densities are different, the hamiltonians on the physical subspace become identical. A hamiltonian
analysis of (\ref{6}) has also been done in \cite{af}. 

The definition of
the canonical momenta  leads to a primary constraint,
\be
\hat\pi_0^a = 0
\label{20}
\ee
and,
\be
\hat\pi_i^a = -\hat F_{0i}^a = \hat E_i^a
\label{21}
\ee
The hamiltonian follows from the Legendre transform and the use of
certain symmetry operations,
\be
\hat H=\int d^3x \,\, Tr\Big({1\over 2}\hat\pi_i\hat\pi_i + {1\over 4}
\hat F_{ij}\hat F_{ij} + \hat{(D_i\pi_i)}\hat A_0\Big)
\label{22}
\ee
where the (hat) covariant derivative is defined as,
\ber
\hat{(D_i\pi_i)}^a &=& \p_i\hat\pi_i^a +i(\hat\pi_i * \hat A_i - \hat A_i *\hat \pi_i)^a \cr
&=& \p_i\hat\pi_i^a + {1\over 2}f^{abc}\{\hat A_i^b,
\hat\pi_i^c\}_*-{i\over 2}d^{abc} [\hat A_i^b,
\hat\pi_i^c]_*
\label{23}
\eer
Here both the commutator and the anticommutator involve the star multiplication. The above
equation defines the Gauss operator, whose vanishing yields the secondary
constraint. As done earlier we pass to the physical sector by imposing
this constraint. In terms of the (hat) electric and magnetic fields, the
hamiltonian reduces to,
\be
\hat H = {1\over 2}\,\,\int d^3x\,\, Tr\Big(\hat {\bf E}^2+\hat
{\bf
B}^2\Big)
\label{24}
\ee
It should be pointed out this is an exact result valid to all orders in
the expansion parameter $\t$. To compare with (\ref{19}), we use the map
(\ref{9}). Then it would be  useful to recast this map in terms of the
electric and magnetic fields,
\be
\hat E_i^a = E_i^a +{1\over 2}\t^{lm}d^{abc}\Big(\e_{imp}B_p^c
E_l^b-A_l^b\p_mE_i^c+{1\over 2}f^{cde}A_l^bA_m^eE_i^d\Big)
\label{25}
\ee

\be
\hat B_i^a = B_i^a -{1\over 4}\t^{lm}d^{abc}\Big(\e^{lmp}B_p^b
B_i^c +2A_l^b\p_mB_i^c- f^{cde}A_l^bA_m^eB_i^d\Big)
\label{26}
\ee
which leads to,
\be
Tr {\bf \hat E}^2= Tr\Big({\bf E}^2
(1-{\bf \t .
B})  + {\bf \t . E} ({\bf B.E} + {\bf E.B})
-{\bf \t} . {\bf \nabla} \times ({\bf E}^2{\bf A})\Big)
\label{27}
\ee

\be
Tr {\bf \hat B}^2= Tr\Big({\bf B}^2
(1-{\bf \t .
B})  
-{\bf \t} . {\bf \nabla} \times ({\bf B}^2{\bf A})\Big)
\label{28}
\ee
Using (\ref{27}) and (\ref{28}) in (\ref{24}),  we see that the hamiltonian density is different from that given in (\ref{19}). However the difference is a boundary term so that the hamiltonians in the two cases agree, as announced
earlier. Also, it should be mentioned that the agreement holds only for the physical hamiltonians, obtained from the canonical hamiltonians by dropping the terms proportional to the Gauss constraint in either description.

\section{Seiberg-Witten map as a canonical transformation}

The relations (\ref{16}) and (\ref{21}) connecting the electric
field with the momenta, together with (\ref{25}), 
provide a map between the momenta in the noncommutative space and the
ordinary one,
\be
\hat \pi_i= \pi_i +{\t^{lm}\over 4}\Big(\{F_{lm}, \pi_i\}-
\{A_{l}, D_m\pi_i +\p_m\pi_i\}\Big)
-{\t^{im}\over 2}\{F_{lm}, \pi_l\}
\label{29}
\ee
which may be regarded as the hamiltonian generalisation of the usual
Seiberg-Witten map (\ref{8}). This result is true for any dimensions. If
we specialise to $3+1$-dimensions, it simplifies to,
\be
\hat \pi_i= \pi_i +{\e^{ijk}\over 4}\{F_{jk}, \t^l\pi_l\}-
{\e^{lmn}\t^{n}\over 4}
\{A_{l}, D_m\pi_i +\p_m\pi_i\}
\label{30}
\ee
while in $2+1$ dimensions, with $\t^{lm}=\e^{lm}\t$,  it is even simpler,
\be
\hat \pi_i= \pi_i -
{\e^{lm}\t\over 4}
\{A_{l}, D_m\pi_i +\p_m\pi_i\}
\label{30a}
\ee

We shall next discuss the  possible interpretation of this map as a canonical transformation. Instead of working with the usual canonical coordinates and momenta, we might as well formulate the discussion in terms of the electric and magnetic fields that have been considered so far.  Also, since all the essential features are contained in the
abelian version itself, henceforth  we confine to this case. It also
simplifies the algebra and permits a quick check with some existing results
\cite{gjp}.

As a first step it is  necessary
to derive the algebra among the electric and magnetic fields.
Using the definition (\ref{16}) of the electric field in terms
of the canonical variables, it is possible to get the complete algebra,
\be
\{B_i(x),  B_j(y)\} =0
\label{40}
\ee

\be
\{E_i(x), B_j(y)\}=-(1+\t^k B_k(x))\Delta_{ij}(x-y)+\t^{\{i}B_{k\}}(x)
\Delta_{kj}(x-y)
\label{41}
\ee

\be
\{E_i(x), E_j(y)\}=\t^{[j}E_{k]}(y)
\Delta_{ik}(x-y) + \t^{[i}E_{k]}(x)
\Delta_{jk}(x-y) +\t^k(E_k(y)-E_k(x))
\Delta_{ij}(x-y)
\label{42}
\ee
where,
\be
A_{\{i}B_{k\}}=A_iB_k+ A_k B_i, \,\, A_{[i}B_{k]}=A_iB_k- A_k B_i
\label{43}
\ee
and the derivative operator has been absorbed in,
\be
\Delta_{ij}(x-y)=\e_{ijk}\p_k\delta(x-y)
\label{44}
\ee

Apart from (\ref{40}), the other two brackets involve correction terms
which vanish with the vanishing of the expansion parameter. In
particular, the electric fields become noncommuting. Moreover, since this
noncommutativity is field dependent, it is essential to verify the Jacobi
identity for the complete algebra. Those involving three $B$ terms are
trivially zero. The Jacobi involving three $E$ terms is also trivial,
since the terms are of $O(\t^2)$. The one involving two $B$ terms and one
$E$ term is also zero, as may be quickly verified by simple inspection.
The only nontrivial check involves two $E$ terms and one $B$ term. Some
amount of algebra is now required. We find,
\be
\{E_i(x), \{E_j(y), B_k(z)\}\}=\t^n\Delta_{in}(x-y)\Delta_{jk}(y-z)
-\Big(\t^j\Delta_{in}(x-y) +\t^n\Delta_{ij}(x-y)\Big)\Delta_{nk}(y-z)
\label{44a}
\ee
The other double bracket is obtained by interchanging $i$ with $j$ and
$x$ with $y$, so that a total of six terms emerges from this algebra. 
There is an exact one to one cancellation of these terms with the six
terms obtained from the final $B-E-E$ double bracket, so that the Jacobi
identity is satisfied.

As an illustration of the use of this involved algebra, we show that it correctly reproduces the lagrangian equations of motion, in the hamiltonian formulation.
The equations of motion obtained from the lagrangian (\ref{12}) can be
expressed in terms of a displacement field ${\bf D}$ and a magnetic field
${\bf H}$ as \cite{gjp},
\be
\nabla . {\bf D} = 0
\label{36}
\ee

\be
{\p\over \p t}{\bf D} - \nabla\times {\bf H}=0
\label{37}
\ee
where,
\be
{\bf D}=(1-{\bf\t} .{\bf B}){\bf E} +({\bf\t} .{\bf E}){\bf B}
+({\bf B} .{\bf E}){\bf \t}
\label{38}
\ee

\be
{\bf H}=(1-{\bf\t} .{\bf B}){\bf B} -({\bf\t} .{\bf E}){\bf E}
+{1\over 2}({\bf E}^2 - {\bf B}^2){\bf \t}
\label{39}
\ee

We now reproduce these equations in the hamiltonian formulation. The
first of these equations is just the Gauss constraint (\ref{18}). This is
easily verified by looking at the definition of the canonical momenta 
(\ref{15}) and identifying it with the displacement field ${\bf D}$.
Effectively (\ref{37}) is the genuine equation of motion since it
involves the accelerations. 
Having obtained the complete algebra, the equations of motion for the
magnetic and electric fields are obtained by bracketing with the
hamiltonian (\ref{19}). It yields,
\be
{\p\over\p t}{\bf B} = - \nabla\times {\bf E}
\label{45}
\ee

\be
{\p\over\p t}{\bf E} =  \nabla\times {\bf H}
+ {\bf M}    
\label{46}
\ee
where,
\be
{\bf M}= \Big({\bf E}.\nabla\times{\bf E}
-{\bf B}.\nabla\times{\bf B}\Big){\bf \t}
+{\bf \t}.{\bf B}(\nabla\times{\bf B})
+{\bf \t}.{\bf E}(\nabla\times{\bf E})
-{\bf \t}.(\nabla\times{\bf
E}){\bf
E}-{\bf \t}.(\nabla\times{\bf B}){\bf B}
\label{47}
\ee
and ${\bf H}$ has been defined in (\ref{39}).

The first of these equations is the standard Maxwell equation,
which is a
consequence of the definition of the electric and magnetic fields in
terms of the field tensor. The second equation gives a complicated time
evolution for the electric field, where the correction terms to the usual
Maxwell equation have been isolated. Note that in the limit of vanishing
$\t$, the ordinary Maxwell equations are reproduced, as expected. It is
now possible to express the additional ${\bf M}$ term in (\ref{46}) as a
total time derivative. To do this the curl of the ${\bf E}$ field is
replaced by the time derivative of the ${\bf B}$ field by using the
identity (\ref{45}). The curl of the ${\bf B}$ is likewise replaced by
the time derivative of the ${\bf E}$ field since any correction to this
Maxwell equation must involve terms of $O(\t)$ leading to terms of
$O(\t^2)$ in ${\bf M}$, that can be ignored. Thus we find,
\be
{\bf M} = {\p\over \p t} \Big(({\bf\t} .{\bf B}){\bf E} -({\bf\t} .{\bf
E}){\bf B} -({\bf E}. {\bf B}){\bf \t}\Big)
\label{48}
\ee

It is now trivial to reproduce (\ref{37}).
 This completes our demonstration of the
equivalence of the equations of motion obtained in the lagrangian and
hamiltonian formulations.

It may be recalled that we had introduced redefined electric $\bar{\bf E}$
(\ref{19a}) and magnetic $\bar{\bf B}$ (\ref{19b}) fields in terms of
which both the lagrangian (\ref{19d}) and hamiltonian (\ref{19c}) assumed
the same structures as in the usual theory. The equations of motion in
these variables, as well as their algebra, can be easily obtained from
our results.

Finally, in order to discuss the role of the Seiberg-Witten map as a canonical transformation, we consider  the issue of the stability of the Poisson algebra
among the electric and magnetic fields under the transformation
(\ref{25}) and (\ref{26}) and the algebra (\ref{40}), (\ref{41}),
(\ref{42}). In the
hat variables, the electric field is the momenta (\ref{21}) conjugate to
the potential
$\hat A_i$, so that,
\be
\{\hat B_i(x),  \hat B_j(y)\} =\{\hat E_i(x),  \hat E_j(y)\}= 0
\label{49}
\ee

\be
\{\hat E_i(x),  \hat B_j(y)\}= -\e_{ijk}\p_k \d(x-y)  +
\e_{ijk}\e_{lmn}\t^n \p_l \hat A_k \p_m\d(x-y)
\label{50}
\ee

Now these brackets are computed in a different way. Using the map
(\ref{25}) and (\ref{26}), the hat variables are expressed in terms of the ordinary variables. The algebra among these variables, given in (\ref{40})- \ref{42}), is then used.  A slightly lengthy algebra leads to the following brackets,
\ber
\{\hat B_i(x),  \hat B_j(y)\} &=& 0\cr
\{\hat E_i(x),  \hat B_j(y)\} &=& -\e_{ijk}\p_k \d(x-y)  +
\e_{ijk}\e_{lmn}\t^n \p_l  A_k \p_m\d(x-y)\cr
\{\hat E_i(x),  \hat E_j(y)\} &=&\e_{ijk}\t^k(\p_l E_l)\d(x-y)
\label{big}
\eer

The algebra among the magnetic fields is trivially identical to (\ref{49}). If we make further use of the Seiberg-Witten map, we see that $\hat A$ in (\ref{50}) can be identified with the usual $A$ in (\ref{big}), since the corrections  will at least involve terms up to $(\t^2)$. Thus the electric-magnetic field bracket in the two cases agree. The last bracket among the electric fields has to be interpreted with some care.  The point is that, on account of (\ref{16}), up to the order we are dealing with, it is possible to rewrite it as,
\be
\{\hat E_i(x),  \hat E_j(y)\}=\e_{ijk}\t^k(\p_l E_l)\d(x-y)=
\e_{ijk}\t^k(\p_l \pi_l)\d(x-y)
\label{51}
\ee

 The bracket is thus proportional to the Gauss
constraint (in the usual variables) , so that on the physical sector, the second equality in
(\ref{49}) is obtained.  This shows that, with a suitable interpretation, the Seiberg-Witten map may be regarded as a canonical transformation.
 
 \section{Application to gauge symmetries}
 
 \bigskip
 
 It is possible to provide an application of the hamiltonian formulation to derive the gauge symmetry of the noncommutative theory governed by the action (\ref{6}). Moreover the analysis is completely general and not confined to any specific order in the expansion parameter $\theta$. It is known by inspection that the action (\ref{6}) is invariant under star gauge transformations, whose infinitesimal version is given by,
 \be
 \d \hat A_\mu^a = \hat{(D_\mu\e)}^a
 \label{n1}
 \ee
 where $\e$ is the gauge parameter. The (hat) covariant derivative is defined in (\ref{23}). We now derive this result.
 
 We adopt the same techniques developed earlier for treating conventional gauge theories
 \cite{ht, brr}. However, there are some subtle issues related to the fact that ordinary multiplication gets replaced by star multiplication. Any gauge theory in the hamiltonian formulation is characterised by the following involutive (first class) algebra involving the constraints $\Phi$ and the canonical hamiltonian $H$,
 \ber
 [H, \Phi_a(x)] &=& \int dy V_a^b(x, y)\Phi_b(y) \cr
 [\Phi_a(x), \Phi_b(y)] &=& \int dz C_{ab}^c(x, y, z)\Phi_c(z)
 \label{n2}
 \eer
 where $V, C$ denote the structure functions which, in the general case, may depend on the phase space variables. The symbols $a, b$ etc. contain the symmetry indices, as well as the number of the constraints. The brackets denote the usual Poisson brackets. Furthermore, in the noncommutative case, the multiplication in the right side is replaced by a star multiplication. However since the expressions are within an integral, the star multiplication can be replaced by the ordinary one and hence (\ref{n2}), as it stands, is correct. Following the Dirac algorithm, the gauge generator $G$ is defined as a linear combination of the first class constraints,
 \be
 G=\int dx \e^a(x)\Phi_a(x)
 \label{n3}
 \ee
 where `$a$' enumerates the constraints. 
 Now  all the gauge parameters $\e^a$ are not independent. The number of independent parameters is identical to the number of primary first class constraints. The other ones are fixed by the following conditions \cite{ht, brr},
 \be
 {d\e^{b_2}(x)\over dt} = \int dy \e^a(y) V_a^{b_2}(y, x) +\int dy dz \e^a(y)v^{a_1}(z)C_{a_1a}^{b_2}(z,y,x)
 \label{n4}
 \ee
 Here the labels 1 and 2 denote the primary and secondary sectors, respectively, so that the full set of constraints denoted by the label  `$a$' would be divided into two parts- $a_1$ denoting the primary constraints (i.e. those obtained from the basic definition of the canonical momenta) and $a_2$ denoting the secondary constraints (i.e. those found by the consistency requirement of time conservation of the primary constraints). The lagrange multipliers entering in the definition of the total hamiltonian $H_T$ are given by $v^{a_1}$,
 \be
 H_T= H +\int v^{a_1}\Phi_{a_1}
 \label{n5}
 \ee
 where $\Phi_{a_1}$ denotes the  primary first class constraints.  In the noncommutative theory, the products in the various integrands are replaced by the star products.  Then the second integral in (\ref{n4}) requires care since it would involve the star product of three objects.  However, in the present problem, this term actually vanishes, as shown below.
 
 We have one primary constraint (\ref{20}) and one secondary (Gauss) constraint (\ref{23}). 
 Following our conventions, we label these constraints as,
 \ber
 \Phi_1 &=& \hat \pi_0\cr
 \Phi_2 &=& \hat{D_i\pi_i}
 \label{n7}
 \eer
The Poisson algebra involving the primary constraint is trivial,
\be
[\Phi_1, \Phi_1] = [\Phi_1, \Phi_2] = 0
\label{no}
\ee
This implies that the $C$-function in (\ref{n4}) vanishes. A non-vanishing piece arises only if the algebra of the primary constraint is non-trivial.
 Thus we obtain,
 \be
 {d\e^{b_2}(x)\over dt} = \int dy \e^a(y) V_a^{b_2}(y, x)
  \label{n6}
 \ee
 
 The variations of the fields are now defined  by bracketing with the generator (\ref{n3}) in the  manner,
 \be
 \d \hat A_\mu(x) = \int dy \e^a(y) * [\hat A_\mu(x), \Phi_a(y)] \,\,\,; \,\, a=1, 2
 \label{n7}
 \ee
  The variation of the time component of the field is easily obtained, with the contribution coming from the primary sector,
 \be
  \d \hat A_0(x) = \int dy \e^1(y) * \d (x-y) =  \int dy \e^1(y)  \d (x-y) = \e^1(x)
  \label{n8}
  \ee
The variation of the space component gets a nonvanishing contribution from the Gauss constraint. This constraint is first written in terms of the structure constants ($f^{abc}, d^{abc}$) of the symmetry group (see (\ref{23})), after which the star products are expanded in full. Computing the algebra and dropping boundary terms, one finally obtains,
\be
\d \hat A_i(x) = \hat{D_i\e^2}(x)
\label{n9}
\ee
However  $\e^1$  and $\e^2$ are not independent. It is possible to determine $\e^1$ in terms of $\e^2$ by using (\ref{n6}). The first step is to obtain the $V$-functions. The algebra of the constraints with the canonical hamiltonian (\ref{22})  is given by,
\ber
[\hat H, \hat\pi_0^a] &=& \hat{(D_i\pi_i)}^a\cr
[\hat H, \hat{(D_i\pi_i)}^a] &=& {1\over 2}f^{abc}\{\hat A^{0b}, \hat{(D_i\pi_i)}^c\}_*-{i\over 2}d^{abc}[\hat A^{0b}, \hat{(D_i\pi_i)}^c]_*
\label{n10}
\eer
where the closure of the Gauss constraint involves both (star) commutator and anticommutator.
While the closure of the primary constraint is trivial, that of the Gauss constraint requires some algebra, but the details are given in \cite{af}.  The $V$-functions appearing in (\ref{n2}) are now obtained from the above algebra. The only nonvanishing ones are given by,
\ber
(V_1^2)^{ab}(x, y) &=& \d^{ab}\d(x-y)\cr
(V_2^2)^{ab}(x, y) &=&  -{1\over 2}f^{abc}\{\d(x-y), \hat A^{0c}(y)\}_*-{i\over 2}d^{abc}[
\d(x-y), \hat A^{0c}(y)]_*
\label{n11}
\eer
In deriving these relations use has been made of the cyclicity of the star product within an integral,
\be
\int\, (A*B*C) = \int\, (B*C*A) = \int\, (C*A*B)
\label{n12}
\ee
and the operator identity,
\be
A(x)*\d(x-y) = \d(x-y)*A(y)
\label{n13}
\ee
There is only one equation for (\ref{n6}) that has to be solved. Writing it out in an expanded form, we find,
\be
 {d\e^{2a}(x)\over dt} = \int dy \e^{1b}(y) (V_1^{2})^{ba}(y, x) +\int dy \e^{2b}(y) (V_2^{2})^{ba}(y, x)
  \label{n14}
\ee
Using the first equation in (\ref{n11}), the first integral in the above relation is trivially done to yield,
\be
\int dy \e^{1b}(y) (V_1^{2})^{ba}(y, x)= \e^{1a}(x)
\label{n100}
\ee
 The second involves the star product of three objects{\footnote{Note that, although not written explicitly, star products are always implied in such products. It is not written explicitly since the star product of two objects within an integral can be replaced by an ordinary product.}}. We evaluate it in some details.
\ber
\int dy \e^{2b}(y) (V_2^{2})^{ba}(y, x) &=& {1\over 2}f^{abc}\int dy \e^{2b}(y)*\Big(\d(x-y)*\hat A^{0c}(x)+\hat A^{0c}(x)*\d(x-y)\Big) \cr
-{i\over 2}d^{abc}\int dy \e^{2b}(y)&*&\Big(
\d(x-y)* \hat A^{0c}(x)-  \hat A^{0c}(x)*\d(x-y)\Big)
\label{n15}
\eer
The star operation is meaningful when the variables are defined at the same point. Thus the argument of the potential has to be converted from $x$ to $y$. This is done by using the identity
(\ref{n13}). Finally, using (\ref{n12}), we get,
\ber
\int dy \e^{2b}(y) (V_2^{2})^{ba}(y, x) &=& {1\over 2}f^{abc}\int dy \d(x-y)*\{\e^{2b}(y), 
\hat A^{0c}(y)\}_* \cr
&-&{i\over 2}d^{abc}\int dy \d(x-y)*[\e^{2b}(y),  \hat A^{0c}(y)]_*
\label{n16}
\eer
Replacing the first star product by an ordinary product, the integrals are evaluated by using the delta function to yield,
\be
\int dy \e^{2b}(y) (V_2^{2})^{ba}(y, x) ={1\over 2}f^{abc}\{\e^{2b}(x), 
\hat A^{0c}(x)\}_* 
-{i\over 2}d^{abc}[\e^{2b}(x),  \hat A^{0c}(x)]_*
\label{n17}
\ee
Using (\ref{n14}), (\ref{n100}) and (\ref{n17}), we obtain,
\ber
\e^{1a}&=&\dot \e^{2a} -{1\over 2}f^{abc}\{\e^{2b}(x), 
\hat A^{0c}(x)\}_* 
+{i\over 2}d^{abc}[\e^{2b}(x),  \hat A^{0c}(x)]_*\cr
&=& \hat{(D_0\e^2)}^a
\label{n18}
\eer
Combining (\ref{n8}), (\ref{n9}) and (\ref{n18}), we obtain the covariant transformation law,
\be
 \d \hat A_\mu^a = \hat{(D_\mu\e^2)}^a
 \label{n19}
 \ee
 which reproduces (\ref{n1}) once $\e^2$ is identified with $\e$. This completes our derivation of the infinitesimal gauge transformations. Note that it has the correct form in the limit $\t\rightarrow 0$, when the noncommutative gauge transformations reduce to the standard nonabelian gauge transformations.
 
 As a further application of the hamiltonian approach, it is possible to prove that the ``Maxwell" type equations (\ref{36}) and (\ref{37}) satisfied by noncommutative electrodynamics in the usual variables are actually valid for any order in $\t$ and not just up to the first order, which has been done explicitly. The first of these equations, as discussed there, is the Gauss constraint, with the displacement field ${\bf{D}}$ identified with the canonical momenta conjugate to ${\bf{A}}$. This constraint is the generator of time independent gauge transformations.
 Since the nature of the gauge transformation $(\d A_\mu=\p_\mu\epsilon)$ generating the abelian gauge symmetry remains the same,  independent of the specific order of $\t$, such an equation characterising the Gauss constraint will always occur. Thus we have,
 \be
 \nabla.{\cal{ D}}=0
 \label{m1}
 \ee
 where,
 \be
 {\cal D}_i= E_i + ..........
 \label{m2}
 \ee
 where the dots represent additional terms in distinct orders of the expansion parameter $\t$. Here ${\cal D}_i$ is the momenta conjugate to $A_i$ in the full $\t$-expanded theory. The leading term is given by the usual electric field since it is the  conjugate momenta in the conventional Maxwell theory. The first correction term in $\t$ has been explicitly given in (\ref{38}).
 
 Having obtained the constraint, the genuine equation of motion can now be derived. This is found by bracketing the ${\cal D}$ with the hamiltonian. Since the electromagnetic field tensor is the local gauge invariant object, the gauge invariant lagrangian will always be constructed in terms of this tensor, as for instance shown in (\ref{12}). Likewise the physical hamiltonian (i.e. the canonical hamiltonian after the imposition of the Gauss constraint) will be constructed in terms of the gauge invariant variables, which are the momenta (${\cal D}_i$) and the magnetic field $B_i$.  The hamiltonian (\ref{19}) has this form once the electric fields are replaced in favour of the momenta by using (\ref{16}). The contribution to the bracket of ${\cal D}_i$ with the hamiltonian will come only from the magnetic field terms. Since the magnetic field is expressed as the curl of the vector potential, the result will involve the curl of something so that one obtains,
 \be
 {\p\over\p t}{\cal D} - \nabla\times {\cal H} = 0
 \label{m3}
 \ee  
 where,
 \be
 {\cal H}_i = B_i + .....
 \label{m4}
 \ee
 As before, the dots denote additional terms in powers of $\t$. The leading term is given by the usual magnetic field to correctly reproduce the standard Maxwell equation, in analogy to the Gauss law, in the limit $\t\rightarrow 0$. The first correction term is given in (\ref{39}).
 
 Equations (\ref{m1}) and (\ref{m3}) are the analogues of (\ref{36}) and (\ref{37}). The fact that the equations can always be put in this form justifies the terminology ``noncommutative electrodynamics". This feature may be used to extend the $O(\t)$ results, as for instance obtained in \cite{gjp, c}. More recently, explicit computations up to $O(\t^2)$ in \cite{bcc}, also confirm the general structure given here.
 
 \section{Discussions}
 
 We have analysed the noncommutative Yang-Mills theory both in the original variables defined in the noncommutative space and also in an effective version by exploiting the Seiberg-Witten map to 
recast the theory in conventional variables.  
Noncommuting
electric fields  are a natural outcome of the effective
theory.  They are essential to correctly reproduce the equations of motion in
the hamiltonian formulation. This noncommutativity  may be interpreted as a sort of
distortion in noncommutating electrodynamics, since the electric fields are now a nonlinear function of the canonical variables. Something similar happens
when the plane waves of the commuting Maxwell theory get distorted in
noncommutating electrodynamics
\cite{gjp, c}, leading to a violation of the superposition principle
\cite{bcc}.  It was possible, after a suitable change of variables, to
express both the lagrangian and hamiltonian in the form of usual gauge
theories, with the complete effect of noncommutativity relegated to the
nontrivial algebra. Although noncommuting electric fields have appeared
earlier-as for instance in topologically massive gauge theories
\cite{djt} or in anomalous gauge theories \cite{rb}-here these are field
dependent, in contrast to the examples cited. 

Our analysis clarified certain issues concerning the reduction of the noncommutative theory to an effective theory by the application of the Seiberg-Witten map. This was possible either through the lagrangian or through the hamiltonian.  As shown in \cite{gkp}, the results need not agree in general. For example, the energy momentum tensor computed from the original action and then expanded by the Seiberg-Witten map is not the same as obtained by first expanding the action itself by this map, after which the tensor is computed. In our analysis, however, we have shown that the hamiltonian, which is a component of the energy momentum tensor, agrees by either approach, provided a passage to the physical subspace is done by a suitable imposition of the appropriate Gauss constraint in the different descriptions. The mismatch in \cite{gkp} comes about presumably because these constraints 
(\ref{18}) and (\ref{23}) do not get mapped by the Seiberg-Witten transformation{\footnote{The fact that the Gauss constraints are not mapped may be seen easily in the $U(1)$ case. In the usual picture, this constraint is gauge invariant, while in the noncommutative picture, it is gauge covariant. 
Since the map connects gauge equivalent classes only, the two constraints do not get identified.}}. Thus it is essential to first pass on to the physical subspace before applying the map. In that case there is no mismatch. Indeed, the imposition of the Gauss law was also crucial for showing that the 
hamiltonian version of the Seiberg-Witten map discussed here, was a canonical transformation.

The hamiltonian formulation was applied to explicitly derive the gauge symmetries, that are otherwise postulated on inspection,  of the 
action in the noncommutative variables. The result was valid for any order in $\t$.  The nontrivial algebra of the first class constraints and the hamiltonian, as well certain properties of field variables and distributions operated under the star multiplication, were essential for getting this result. Furthermore, based on the structure of the gauge generators it was possible to show that the equations of motion for the $\t$-expanded noncommutative electrodynamics could always be cast in the form of ``Maxwell" equations. This suggests that results, which have been computed up to the first order in $\t$, could possibly be extended for higher orders.

The implications of our analysis at the quantum level may be discussed both in the effective ($\t$)-expanded theory and in the original noncommutating variables. In the former case, the Poisson brackets among the gauge invariant variables obtained here in (\ref{40}), (\ref{41}) and (\ref{42}) can be expressed as commutators since the expressions involve only linear variables so that there is no ordering ambiguity.  Expressions involving products of field variables should be defined by the Weyl ordering. This is the natural  prescription once it is realised that the star 
operation, through the composition law, directly leads to this ordering. In the noncommutating variables, the basic brackets are just the usual Poisson brackets. Complications arise because of the nontrivial algebra of the constraints and the hamiltonian. However we have shown here how to handle such complications, precisely obtaining the gauge symmetries. The construction of the BRST charge, which acts as the generator of the BRST transformations, should be possible 
along these lines, with suitable inclusion of ghosts, in analogue to usual Yang-Mills theory. 
Another aspect is the well known fact that the process of obtaining a quantum theory from a classical one by constraining and quantising is not commutative; i.e. one can get different results by first solving the constraints and then quantising or by reversing the process. Here we have found that the hamiltonian in the $\t$-expanded theory and the one in the original variables get mapped only if the constraints are first eliminated. Otherwise we get two distinct formulations. A more detailed description of these and other issues related to the quantisation procedure are beyond the scope of the present paper. 

We conclude by mentioning that, noncommutative field theory, being an emergent topic, it is desirable to explicitly show that the hamiltonian formulation is consistent with the lagrangian one, even at the classical level. Indeed certain discrepancies were already mentioned \cite{gkp}. We have shown how to obtain a consistent formulation, apart from providing certain applications of the  hamiltonian approach that could be used for a quantum analysis. Finally, since the Seiberg-Witten map connects gauge equivalent classes, it is expected to be related to canonical transformations. We proved this explicitly in the physical subspace, using the hamiltonian analysis.

\bigskip

{\Large{\bf Acknowledgments}}

I thank the Japan Society for Promotion of Science (JSPS) for 
support and the members of the theory group, KEK, for their gracious
hospitality.

\end{document}